
\magnification=1200


\def\gl{\mathrel{\raise1ex\hbox{$>$\kern-.75em\lower1ex\hbox{$<$}}}}
\def\lg{\mathrel{\raise1ex\hbox{$<$\kern-.75em\lower1ex\hbox{$>$}}}}
\def\gtwid{\mathrel{\raise.3ex\hbox{$>$\kern-.75em\lower1ex\hbox{$\sim$}}}}
\def\ltwid{\mathrel{\raise.3ex\hbox{$<$\kern-.75em\lower1ex\hbox{$\sim$}}}}
\def\sqr#1#2{{\vcenter{\hrule height.#2pt
      \hbox{\vrule width.#2pt height#1pt \kern#1pt
         \vrule width.#2pt}
      \hrule height.#2pt}}}

\overfullrule=0pt

\def\eg{\hbox{{\it e.\ g.}}}\def\ie{\hbox{{\it i.\ e.}}}
\def\vv{{\it vice versa}}


\def\leaderfill{\leaders\hbox to 1em{\hss.\hss}\hfill}


\def\CN{\hbox{{$\cal N$}}}

\def\CP{\hbox{{$\cal P$}}}

\def\ref#1{${}^{#1}$}
\newcount\eqnum \eqnum=0  
\newcount\eqnA\eqnA=0\newcount\eqnB\eqnB=0\newcount\eqnC\eqnC=0\newcount\eqnD\eqnD=0
\def\eqnoi{\global\advance\eqnum by 1\eqno(\the\eqnum)}
\def\eqnai{\global\advance\eqnum by 1\eqno(\the\eqnum{a})}
\def\eqnbi{\eqno(\the\eqnum{b})}
\def\eqnci{\eqno(\the\eqnum{b})}
\def\eqnoA{\global\advance\eqnA by 1\eqno(A\the\eqnA)}
\def\eqnoB{\global\advance\eqnB by 1\eqno(B\the\eqnB)}
\def\eqnoC{\global\advance\eqnC by 1\eqno(C\the\eqnC)}
\def\eqnoD{\global\advance\eqnD by 1\eqno(D\the\eqnD)}
\def\back#1{{\advance\eqnum by-#1 Eq.~(\the\eqnum)}}
\def\backs#1{{\advance\eqnum by-#1 Eqs.~(\the\eqnum)}}
\def\backn#1{{\advance\eqnum by-#1 (\the\eqnum)}}
\def\backA#1{{\advance\eqnA by-#1 Eq.~(A\the\eqnA)}}
\def\backB#1{{\advance\eqnB by-#1 Eq.~(B\the\eqnB)}}
\def\backC#1{{\advance\eqnC by-#1 Eq.~(C\the\eqnC)}}
\def\backD#1{{\advance\eqnD by-#1 Eq.~(D\the\eqnD)}}
\def\last{{Eq.~(\the\eqnum)}}                   
\def\lasts{{Eqs.~(\the\eqnum)}}                   
\def\lastn{{(\the\eqnum)}}                      
\def\lastA{{Eq.~(A\the\eqnA)}}\def\lastB{{Eq.~(B\the\eqnB)}}
\def\lastC{{Eq.~(C\the\eqnC)}}\def\lastD{{Eq.~(D\the\eqnD)}}
\newcount\refnum\refnum=0  
\def\refi{\smallskip\global\advance\refnum by 1\item{\the\refnum.}}

\newcount\rfignum\rfignum=0  
\def\rfigi{\medskip\global\advance\rfignum by 1\item{Figure \the\rfignum.}}

\newcount\fignum\fignum=0  
\def\figi{\global\advance\fignum by 1 Fig.~\the\fignum}

\newcount\rtabnum\rtabnum=0  
\def\rtabi{\medskip\global\advance\rtabnum by 1\item{Table \the\rtabnum.}}

\newcount\tabnum\tabnum=0  
\def\tabi{\global\advance\tabnum by 1 Table~\the\tabnum}

\newcount\secnum\secnum=0 
\def\chap#1{\global\advance\secnum by 1
\bigskip\centerline{\bf{\the\secnum}. #1}\smallskip\noindent}

\newcount\nlet\nlet=0
\def\numblet{\relax \global\advance\nlet by 1
\ifcase \nlet \ \or a\or b\or c\or d\or e\or
f\or g\or h\or i\or j\or k\or l\or m\or n\or o\or p
\or q \or r \or s \or t \or u \or v \or w \or x \or y
\or z \else .\nlet \fi}
\def\asubsec#1{\smallskip{\bf\centerline{(\numblet) #1}\smallskip}}

\newcount\rlet\rlet=0
\def\romlet{\relax \global\advance\rlet by 1
\ifcase \rlet \ \or I\or II\or III\or IV\or V\or
VI\or VII\or VIII\or IX\or X\or XI\or XII\or XIII\or XIV\or XV\or XVI
\or XVII \else .\rlet \fi}
\def\rchap#1{\bigskip\centerline{\bf{\romlet}. #1}\smallskip\noindent}

\def\p2d#1#2{{\partial^2 #1\over\partial #2^2}} 
\def\t2d#1#2{{d^2 #1\over d #2^2}} 
\def\av#1{\langle #1\rangle}                    


\def\2kth{{$2k^{\rm th}$}}

\def\n-th{{$(n-1)^{\rm th}$}}

\def\N-th{{$(N-1)^{\rm th}$}}

\def\0th{$0^{\rm th}$}
\def\1st{$1^{\rm st}$}
\def\2nd{$2^{\rm nd}$}
\def\3rd{$3^{\rm rd}$}
\def\4th{$4^{\rm th}$}
\def\5th{$5^{\rm th}$}
\def\5th{$6^{\rm th}$}
\def\6th{$7^{\rm th}$}
\def\7th{$7^{\rm th}$}
\def\8th{$8^{\rm th}$}
\def\9th{$9^{\rm th}$}

\def\a{{\alpha}}
\def\b{{\beta}}

\def\n{\nu}
\def\p{\pi}

\def\t{\tau}

\def\addr{
    \centerline{Center for Polymer Studies and Department of Physics}
     \centerline{Boston University, Boston, MA~ 02215}}


\def\aap #1 #2 #3 {{\sl Adv.\ Appl.\ Prob.} {\bf #1}, #2 (#3)}
\def\acp #1 #2 #3 {{\sl Adv.\ Chem.\ Phys.} {\bf #1}, #2 (#3)}
\def\jpa #1 #2 #3 {{\sl J. Phys.\ A} {\bf #1}, #2 (#3)}
\def\jsp #1 #2 #3 {{\sl J. Stat.\ Phys.} {\bf #1}, #2 (#3)}
\def\pA #1 #2 #3 {{\sl Physica A} {\bf #1}, #2 (#3)}
\def\pra #1 #2 #3 {{\sl Phys.\ Rev.\ A} {\bf #1}, #2 (#3)}
\def\prb #1 #2 #3 {{\sl Phys.\ Rev.\ B} {\bf #1}, #2 (#3)}
\def\pre #1 #2 #3 {{\sl Phys.\ Rev.\ E} {\bf #1}, #2 (#3)}
\def\prept #1 #2 #3 {{\sl Phys.\ Repts.} {\bf #1}, #2 (#3)}
\def\prl #1 #2 #3 {{\sl Phys.\ Rev.\ Lett.} {\bf #1}, #2 (#3)}
\def\prsl #1 #2 #3 {{\sl Proc.\ Roy.\ Soc.\ London Ser. A} {\bf #1}, #2 (#3)}
\def\SIAMjam #1 #2 #3 {{\sl SIAM J.\ Appl.\ Math.} {\bf #1}, #2 (#3)}
\def\wrr #1 #2 #3 {{\sl Water Resources Res.} {\bf #1}, #2 (#3)}
\def\zw #1 #2 #3 {{\sl Z. Wahrsch.\ verw.\ Gebiete} {\bf #1}, #2 (#3)}

\def\pe{P\!e\,}


\centerline{\bf Life and Death  at the Edge of a Windy Cliff}
\vskip 0.25in
\centerline{S.~Redner and P.~L.~Krapivsky}
\bigskip
\addr
\vskip 0.4in
\baselineskip=19 truebp
\centerline{\bf Abstract}\medskip

{
\narrower\narrower

\noindent  The survival probability of a particle diffusing in the two
dimensional domain $x>0$ near a ``windy cliff'' at $x=0$ is
investigated.  The particle dies upon reaching the edge of the cliff.
In addition to diffusion, the particle is influenced by a steady ``wind
shear'' with velocity $\vec v(x,y)=v\,{\rm sign}(y)\,\hat x$, \ie, no
average bias either toward or away from the cliff. For this
semi-infinite system, the particle survival probability decays with time
as $t^{-1/4}$, compared to $t^{-1/2}$ in the absence of wind.  Scaling
descriptions are developed to elucidate this behavior, as well as the
survival probability within a semi-infinite strip of finite width
$|y|<w$ with particle absorption at $x=0$.  The behavior in the strip
geometry can be described in terms of Taylor diffusion, an approach
which accounts for the crossover to the $t^{-1/4}$ decay when the width
of the strip diverges.  Supporting numerical simulations of our
analytical results are presented.

}

\vfill\eject
\baselineskip=22 truebp
\rchap{Introduction}

Consider a particle which diffuses in the  semi-infinite planar
domain $(x>0,y)$ and is absorbed when $x=0$ is reached.  The line $x=0$ can
be viewed as a ``cliff'' and  absorption at $x=0$
corresponds to the particle falling to its death.
For this system, it is well known that the particle survival
probability, $S(t)$, decays in time as [1]
$$
S(t)\sim {x_0\over\sqrt{Dt}}\,\,.\eqnoi
$$
Here $x_0$ is the initial distance from the particle to the cliff and
$D$ is the diffusion coefficient.  In this article, we are interested in
understanding the time dependence of $S(t)$ when the diffusing particle
also experiences a ``wind shear'', defined as the velocity field $\vec
v(x,y)=v\hat x$ for $y>0$ and $\vec v(x,y)=-v\hat x$ for $y<0$ (Fig.~1).
Our primary result is that, although there is no average bias either
toward or away from the cliff, the survival probability decays as
$t^{-1/4}$, compared to the $t^{-1/2}$ decay in the absence of the bias.
This result is contrary to the naive intuition of a faster decay, as
wind shear enhances longitudinal ($x$) diffusion, which, from \last,
should reduce the survival probability.  More generally, our interest is
in understanding the interplay between macroscopically heterogeneous
convection and diffusion on first-passage phenomena.  The wind shear
geometry is a relatively simple example of such a system.  In spite of
this simplicity, relatively unusual first-passage characteristics occur.

Our work is also complementary to a recent paper by Lee and Koplik [2]
where the survival probability of a particle in the same wind shear is
considered, but in a ``two-layer'' system with $y$ unbounded, and with
absorbing boundaries at $x=0$ and at $x=L$.  For this system, Lee and
Koplik gave asymptotic arguments to show that the survival probability
decays as $\exp(-t/T)$ with $T\propto L$.  In comparison, for pure
diffusion in the same domain, $S(t)\sim\exp(-t/T_D)$, but with a decay
time $T_D\propto L^2$.  Thus, in accord with intuition, wind shear
decreases the survival probability in the two-layer system (although the
functional form of $S(t)$ is unchanged).  However, for the semi-infinite
planar system, the wind shear has the opposite effect of enhancing the
particle survival probability and moreover changes the decay exponent
from 1/2 to 1/4.

A fundamental aspect of diffusive motion in wind shear is that the
particle spends relatively long periods of time exclusively in the
region $y>0$ or $y<0$ before returning to $y=0$.  In fact, the
distribution of time intervals between successive crossings of $y=0$
coincides with the first passage probability for a one-dimensional
random walk to return to its starting point.  Since this probability
decays as $t^{-3/2}$, the particle has relatively long alternating
flights where $x$ increases or decreases linearly with time.  For a
particle which starts at the origin in an unbounded two-dimensional
system, this structure for the longitudinal steps leads to the following
unusual probability distribution in $x$ at time $t$ [3,4]:
$$
\CP(x,t)\equiv\int\,dy\,P(x,y,t)\propto {1\over\sqrt{(vt)^2-x^2}}.\eqnoi
$$
Rather surprisingly, the probability distribution is peaked as $x\to\pm
vt$ and is a minimum for $x=0$.  These unusual features are one
manifestation of the classical arcsine law for long leads in a
one-dimensional random walk [3].  The interplay between these long
excursions and the absorbing boundary are responsible for many of the
intriguing features of the particle survival probability.

In section II, we define microscopic lattice rules to model a particle
moving in a wind shear.  In section III, we present scaling arguments to
determine the long-time behavior of $S(t)$ for the semi-infinite planar
system.  Somewhat different behavior occurs if the underlying diffusion
is only in the $y$ (transverse) direction compared to the underlying
diffusion being isotropic.  While the former case is in some sense
simpler, it does not have a smooth limit as bias vanishes.  To
understand this basic limit, we therefore introduce a variant of the
original model which reduces to isotropic diffusion when the bias
vanishes.  In section IV, we apply the Taylor diffusion description to
determine the survival probability in a finite width strip $x>0$ and
$|y|\leq w$, with reflection at $|y|=w$ and absorption at $x=0$.  Our
results for the strip are combined with crossover arguments to provide
additional insights into the corresponding first-passage properties of
the planar system.  A salient feature of the survival probability in the
strip geometry is the non-trivial dependence of the survival probability
on the initial distance from the particle to the cliff, the velocity,
and the diffusion coefficient.  We conclude and discuss some open
questions in section V.

\bigskip
\rchap{Wind Shear in Two Dimensions}

Our system is a square lattice with $x\ge 0$ and $y=0$ defined to bisect
the bonds which join the bottom row in the upper half plane to the top
row in the lower half plane (Fig.~1).  With this construction, there are
no sites with $y=0$, and ambiguities associated with assigning hopping
rates from these sites are avoided.  The hopping rates from any site in
the upper (lower) half plane are spatially homogeneous.  We consider two
different hopping rules which correspond, respectively, to anisotropy
and isotropy (for $v\to 0$) in the underlying diffusive motion.

For anisotropic diffusion, to account for a variable transverse
diffusivity and longitudinal velocity, we define the following hopping
rates to six neighbors of a given site.  For sites in the upper-half
plane:
$$
\eqalign{
p_{1,\pm1}&= {1\over 4}(1+v)\,D_y,          \cr
p_{1,0}   &= {1\over 2}(1+v)(1-D_y),        \cr
p_{-1,\pm1}&={1\over 4}(1-v)\,D_y,          \cr
p_{-1,0}   &={1\over 2}(1-v)(1-D_y).         \cr}\qquad\qquad
{\rm anisotropic~ diffusion}\eqnai
$$
Here the subscript on $p_{\vec r}$ indicates vector displacement defined
by the hop.  In the lower half plane, similar hopping rates exist,
except with an opposite sign for $v$.  With these rules, there is a
diffusivity in the $x$-direction whose magnitude depends on $v$ and
vanishes as $v\to 1$.  For the case of unit velocity and unit transverse
diffusivity, these hopping rules reduce to a 2-site neighborhood.  This
appears to be the simplest implementation for random walk modeling of
diffusion in a wind shear.

While the above hopping rule is suitable for most of our purposes, it
leads to pathological behavior for $v\to 0$ or for $D_y\to 0$.  Because
useful insights can be gained by considering the crossover to purely
diffusive behavior as $v\to 0$, we therefore define a second model in
which the underlying diffusion is isotropic.  For sites in
the upper half plane, we define the hopping rates:
$$
\eqalign{
p_{0,\pm1}  &= D,       \cr
p_{1,0}     &= D+{v\over 2},  \cr
p_{-1,0}    &= D-{v\over 2},  \cr
p_{0,0}     &= 1-4D.       \cr}\qquad\qquad\qquad
{\rm isotropic~ diffusion}\eqnbi
$$
Clearly, these rules can be applied only for $D<1/4$ and $v<1/2$ to
ensure positivity of all the hopping rates.

In the following section, we outline the asymptotic behavior of the
survival probability with these two lattice realizations of wind
shear.

\bigskip
\rchap{Survival Probability in the Planar System}

\asubsec{Wind Shear with Anisotropic Diffusion}

Consider now the survival probability of a particle initially at
$(x_0,0)$ in the semi-infinite planar system $x>0$, with absorption at
$x=0$.  From numerical exact enumeration of the probability distribution
for the case $v=D_y=1$, it is evident that the survival probability
decays as $t^{-1/4}$ (Fig.~2).  A relatively simple way to understand
this result is to focus on those points where the particle trajectory
crosses from $y>0$ to $y<0$ or \vv.  Since the transverse motion is a
one-dimensional random walk, the probability distribution of times
between successive crossings asymptotically varies as $t^{-3/2}$ [1].
Consequently, the longitudinal displacement $x(t)$ versus $t$ is a
L\'evy flight which consists of ``segments'' $t_i$ whose lengths are
distributed according to the above distribution (Fig.~1) [5].  Here, the
term ``segment'' refers to a connected portion of the trajectory with
the $y$-coordinate having the same sign.

The existence of a finite observation time $t$, however, implies that
the segment length distribution is necessarily cut off at this time.
With this cutoff, the average segment length is given by
$\av{t}\propto\int^t\,t'\cdot t'^{-3/2}\,dt'\propto t^{1/2}$.  Thus, as
might be anticipated, a trajectory of $t$ steps can typically be
decomposed into $\CN\propto\sqrt{t}$ segments, each of length
$\sqrt{t}$.  At the segment level, the probability that the walk does
not reach $x=0$ is equivalent to particle survival.  Since there are
$\CN$ independent segments, this no return probability should therefore
vary as $1/\sqrt{\CN}$ which, in turn, is proportional to $t^{-1/4}$, in
accord with our observations.

It is also instructive to consider the full dependence of $S(t)$ on
microscopic parameters.  For our system, $S(t)$ can only be a function
of the basic variables $x_0$, $v$, $D_y$, and $t$.  Since $S(t)$ is
dimensionless, it is convenient to introduce the following groupings
with units of time, $\tau_{||}=x_0/v$ and $\tau_D=x_0^2/D_y$, and write
for the survival probability
$$
S(t)\sim f\bigl({t\over\tau_{||}},{t\over\tau_{D}}\bigr).
\eqnoi
$$
A crucial fact is that the particle survival probability is governed by
the {\it difference} in residence times within the regions $y>0$ and
$y<0$.  As written in the original arcsine law (Eq.~(2)), this
difference is {\it independent\/} of $D_y$.  Combining this with the
fact that $S\propto t^{-1/4}$, the asymptotic form of $S(t)$ reduces to
$$
S(t)\propto \left({\t_{||}\over t}\right)^{1/4} =\left({x_0\over
vt}\right)^{1/4}.\eqnoi
$$
To justify this rather unexpected behavior, it is instructive to
consider the survival probability on a finite width strip (Sec.~IV).
For a system with finite extent $|y|\leq w$, then $w$ and $D_y$
naturally appear in the combination $\t_\perp=w^2/D_y$.  Since any
finite value of $D_y$ leads to the same value of $\t_\perp$ as
$w\to\infty$, it suggests that \last\ should be independent of $D_y$.
Our numerical data supports this conclusion (Fig.~2), as long as
$\t_{||}>1/D_y$, so that there is mixing between the upper and lower
half-planes, before significant absorption occurs.  However, our data
typically exhibits a very small and unexplained residual systematic
dependence of $S(t)$ on $D_y$.

Another interesting aspect of the surviving particles is their spatial
distribution (Fig.~3).  Numerically, the surviving particles are
predominantly within a region whose mean longitudinal and transverse
positions are given by $\av{x(t)}\propto t$ and $\av{y(t)}\propto
t^{1/2}$.  These results can be justified heuristically.  For walks to
survive, there clearly must be a longer residence time in the upper half
plane than the lower half plane.  However, starting from $(x,0)$, a walk
can tolerate a short excursion into the lower half plane and still
survive.  If this excursion extends to a transverse distance
$y\cong-\sqrt{Dx/v}$, then the time required for the particle to return
to ``safety'' ($y>0$) becomes of order of the time for the particle to
be convected to the cliff.  Thus if the particle enters the region
$y<-\sqrt{Dx/v}$, it is likely to be absorbed (Fig.~1).  Since surviving
particles must almost always be in the region $y>-\sqrt{Dx/v}$, and
hence predominantly in the upper half plane, this leads to the
longitudinal displacement of the survivors being proportional to $t$.
In a similar spirit, if the survivors are mostly in the upper half
plane, then their transverse dispersion can be governed only by
diffusion, so that $\av{y(t)}\propto t^{1/2}$.

It is also worth noting that if $\av{x(t)}\propto t$, then the typical
value of $y$ for which absorption occurs is $-\sqrt{Dt/v}$.  Thus the
two-dimensional system can be reduced to an effective one-dimensional
transverse problem of a purely diffusing particle starting in the domain
$y>0$ with an absorbing boundary at $-\sqrt{Dt/v}$.  This latter system
has been extensively studied [6].  It is known that $S(t)\propto
t^{-\a}$ with $\a$ dependent on $D$ and $v$ in such a way so that $\a\to
1/2$ for $v\to 0$ and $\a\to 0$ for $v\to\infty$.  This connection
between the wind shear and one-dimensional moving boundary problems
provides further evidence of a decay exponent for $S(t)$ in wind shear
which is less that 1/2.

\asubsec{Wind Shear with Isotropic Diffusion}

When the underlying diffusion is isotropic, the survival probability
again decays as $t^{-1/4}$ in the long time limit (Fig.~4(a)).  As might be
anticipated, the effect of diffusion is subdominant with respect to
convection in governing the value of exponent in the time dependence of
$S(t)$.  However, for the isotropic system there is a crossover from
diffusive behavior, for small $v$, to convective behavior, as $v$
becomes large.  By applying scaling to determine the nature of this
crossover, we also determine the dependence of $S(t)$ on the physical
parameters of the system.  This aspect merits emphasis, as there are
important differences in the dependence of $S(t)$ on system parameters
with isotropic and anisotropic diffusion.

For isotropic diffusion, the basic time scales of the system are
$$
\tau_D={x_0^2\over D},\quad \tau_{||}={x_0\over v},\quad {\rm and\ \ }
\tau_\times=D/v^2.\eqnai
$$
These times are, respectively, the characteristic time to diffuse to the
cliff, the time to convect to the cliff, and the crossover time beyond
which convection dominates over
diffusion.  It is convenient to measure these times in units of
$\t_{||}$ to give
$$
\tilde\tau_D=\pe,\quad \tilde\tau_{||}=1,\quad {\rm and\ \ }
\tilde\tau_\times={1\over \pe},\eqnbi
$$
where $\pe=x_0v/D$ is the P\'eclet number.  In the case where $\pe>1$,
convection becomes established before the particle can reach the cliff
and the behavior is the same as that discussed previously for
anisotropic diffusion.  On the other hand, for $\pe<1$, the fundamental
times obey the inequalities $\tilde\t_D<\tilde\t_{||}<\tilde\t_\times$.
Consequently, particle death at early times is determined by diffusion
and it is only for $t>\tilde\t_\times$ that convective behavior sets in
(Fig.~4(b)).  In dimensionless units, the early-time decay has the form
$S(t)\sim (\pe/\tilde\t)^{1/2}$, while at late times the survival
probability can be written as $S(\t\to\infty)\sim
A(\pe)\,\tilde\t^{-1/4}$.  Matching these two asymptotes at
$\tilde\t_\times=1/\pe$ fixes the amplitude $A(\pe)$, from which
$$
S(t\to\infty)\propto {(\pe)^{3/4}\over{\tilde\tau^{1/4}}} =
{x_0\,v^{1/2}\over{D^{3/4}\,
t^{1/4}}}.
\eqnoi
$$
As shown in Fig.~4(a), the long-time data accords with scaling behavior
predicted by \last.

\rchap{Survival in the Semi-Infinite Strip}

It is instructive to examine the survival probability in the presence of
wind shear in a semi-infinite strip of width $|y|\leq w$.  Because this
system is effectively one dimensional, the longitudinal motion
asymptotically reduces to diffusion in one dimension.  However, this
motion is properly described by Taylor diffusion which accounts for the
interplay between diffusion and convection [7].  From this description,
the form of the survival probability is, in principle, straightforward
to deduce.  It is then possible to infer properties of the survival
probability in the planar system by letting $w\to\infty$.  However,
there is an unexpected subtlety in the behavior of $S(t)$ for the strip
which depends on the relation between $x_0$ and $w$.  The resolution of
this feature provides useful general insights about the nature of the
survival probability.

Taylor diffusion arises because the particle convects to the right while
$y>0$ and convects to the left when $y<0$.  Since this switching between
$y>0$ and $y<0$ is governed by diffusion, the longitudinal motion is
also diffusive.  The characteristic time to switch between right-going
and left-going segment is the same as the time needed to diffuse between
the regions $y>0$ and $y<0$ (or \vv), namely, $\tau_{\perp}\sim w^2/D$.
Therefore the step length of the effective longitudinal random walk is
$\ell\sim v\t_\perp$ and the corresponding Taylor diffusivity is given
by $D_{||}\sim\ell^2/\t_\perp \sim v^2w^2/D$ [7].

With this characterization of the longitudinal motion, we now
investigate the survival probability in the strip in terms of the two
basic time scales $\t_{||}$ and $\t_{\perp}$.  First consider the limit
$\tau_{||}\gg\tau_{\perp}$, which can be re-expressed as $x_0\gg\ell$.
Physically, many longitudinal segments are needed before the cliff is
reached, or equivalently, the walk encounters the sides of the strip
many times before reaching $x=0$.  Therefore, before any absorption
occurs, the trajectory has time to become truly one-dimensional.
Thus we conclude that the particle survival probability is given by the
suitably adapted one-dimensional expression,
$$
S(t;x_0\gg\ell)\sim {x_0\over\sqrt{D_{||}t}}.\eqnai
$$
In terms of the basic time scales,
this (dimensionless) survival probability can be re-expressed as
$$
S(t;x_0\gg\ell)\sim {\tau_{||}\over\sqrt{\t_{\perp}\,t}}.\eqnbi
$$

However, the converse limit $\tau_{||}\ll\tau_{\perp}$ ($x_0\ll\ell$),
is more relevant for understanding the survival probability in the
planar system with wind shear.  In this case, there is significant loss
of probability by absorption at $x=0$ before the sides of the strip play
any role.  Consequently, the survival probability should decay as
$(x_0/vt)^{1/4}$ at short times, as in the semi-infinite planar system,
and cross over to a decay of the form $Bt^{-1/2}$ when
$t\gg\tau_{\perp}$ (Fig.~5).  By matching these two asymptotes at
$t=\tau_{\perp}$ we determine the amplitude $B$ and thereby find for the
survival probability,

$$
S(t;x_0\ll\ell)\propto \left(\tau_{||}\tau_{\perp}\over t^2\right)^{1/4}.\eqnoi
$$

A noteworthy feature of \back1\ and \last\ is the opposite dependences
in $w$.  While $S(t;x_0\gg\ell)\propto 1/w$, $S(t;x_0\ll\ell)\propto
w^{1/2}$ (and is also proportional to $x_0^{1/4}$).  In the latter case,
$S(t;x_0\ll\ell)$ must be increasing in $w$ for there to be a slower
time dependence in $S(t;x_0\ll\ell)$ when $w=\infty$.  The changeover
from \back1\ to \last\ as $x_0$ becomes smaller than $\ell$ can be
viewed as arising by $x_0$ ``sticking'' at $\ell$.  This occurs because
after one Taylor diffusion step, the walk is either absorbed or is
reintroduced at a distance of order $\ell$ from the absorber.  In the
time $\t_{\perp}\sim w^2/D$ required for this reintroduction, only a
fraction $(x_0/v\t_\perp)^{1/4}$ of the walks remain.  Consequently,
$$
\eqalign{
S(t;x_0\ll\ell)&\cong S(t;x_0\gg\ell)|_{x_0=\ell}\times \left({x_0\over
v\t_\perp}\right)^{1/4}\cr
     &\propto {x_0^{1/4}\,w^{1/2}\over{(Dv)^{1/4}}}\cr}
\eqnoi
$$
Notice also that the limiting expressions for $S(t)$ for $x_0\gg\ell$
and $x_0\ll\ell$ both become of order $w/\sqrt{Dt}$ when $x_0\cong\ell$.
This provides a useful self-consistency check on our approach.

\rchap{Discussion}

We have investigated the behavior of the survival probability for a
diffusing particle with a planar absorbing boundary, or cliff, in the
presence of a superimposed ``wind shear''.  Although this velocity field
does not have any bias either toward or away from the cliff, the
long-time behavior of the survival probability $S(t)$ is strongly
affected by the wind shear.  Our primary result is that $S(t)\propto
t^{-1/4}$ for a semi-infinite plane, compared to a decay of $t^{-1/2}$
for the survival probability in the absence of bias.  Although our
approaches are neither rigorous nor microscopic, they provide a good
quantitative account of simulation results.  It is worth noting that the
image method, which yields the survival probability in the presence of a
planar absorber for both diffusion and and uniform convection, does not
appear to be generalizable to wind shear.

Our prediction for the survival probability has applicability beyond the
case of wind shear.  Consider, for example, a particle diffusing with
superimposed linear shear, $\vec v(x,y)=v\,y\,\hat x$, and with
absorption at $x=0$.  For this system, exact enumeration of the
probability distribution shows that the survival probability decays as
$t^{-1/4}$ [8].  This result is also supported by the segment argument
given in Sec.~III(a) which suggests that $S(t)\propto
(\tau_{||}/t)^{1/4}$, where $\tau_{||}$, now the typical time to convect
from $x_0$ to the cliff in shear flow, is determined by
$v\sqrt{D\tau_{||}}\,\tau_{||}=x_0$.  Thus for the linear shear,
we predict the following asymptotic dependence of the survival
probability on model parameters,
$$
S(t)\propto \left(x_0D\over v\right)^{1/6}(Dt)^{-1/4}.
\eqnoi
$$
In fact, we believe that the $t^{-1/4}$ decay should hold for
any velocity field with $v_x(y)=-v_x(-y)$.  For example,
for power-law shear $\vec v(x,y)=v\,y\,|y|^{\beta -1}\,\hat x$,
the asymptotic form
$$
S(t)\propto \left(x_0D\over v\right)^{1\over 2(\beta +2)}(Dt)^{-1/4}
\eqnoi
$$
is expected.  Notice, that only in the case of wind shear ($\b=0$)
$S(t)$ is independent of $D$.  Another situation for which unusual
behavior of $S(t)$ can be anticipated is stratified random flow [9] in
which $v_x(y)$ is a random zero-mean function of $y$.  In this case,
there could be different behavior for $S(t)$ in a typical configuration
of the velocity field and when averaged over all configurations of
velocities.

It would also be desirable to develop more rigorous and microscopic
approaches to understand the first-passage characteristics of systems
with various types of neutral bias fields.  For the case of wind shear,
the distinct nature of the problem for $y>0$ and $y<0$ suggests that the
Wiener-Hopf technique may be suitable.  In fact, this technique has been
successfully applied to a related problem involving first passage in the
presence of colored noise [10].  However, we have been unable to apply
this method to our problem.

\bigskip
\rchap{Acknowledgments}

We thank A.~Acrivos, D. ben-Avraham, E.~Ben-Naim, C.~R.~Doering,
A.~Drory, J.~Koplik, and J.~Lee for helpful discussions and E.~Ben-Naim
for providing the simulation results for the survival probability in
linear shear flow.  We also gratefully acknowledge NSF grant DMR-9219845
for partial support of this research.


\vfill\eject
\centerline{\bf References}\smallskip

\refi See, \eg, G. H. Weiss and R. J. Rubin, \acp 52 363 1983 , and references
therein.

\refi J. Lee and J. Koplik, \jsp 79 895 1995 .

\refi W.~Feller, {\sl An Introduction to Probability Theory and Its
      Applications} (Wiley, New York, 1968).

\refi E. Ben-Naim, S. Redner, and D. ben-Avraham, \pra 45 7207 1992 .

\refi In the context of Levy flights, the survival probability has been
considered using different approaches in G.~Zumofen and Y.~Klafter, \pre
51 2805 1995 .

\refi See, \eg, L. Breiman, in {\sl Proc.\ Fifth Berkeley Symposium
Math.\ Statist.\ and Probab.}, vol.\ II (1967); K. Uchiyama, \zw
54 75 1980 ; P. Salminen, \aap 20 411 1988 ; L. Turban \jpa
25 L127 1992 ; F. Igl\'oi, \pra 45 7024 1992 .  A pedagogical
account is given in P.~L.~Krapivsky and S.~Redner, preprint.

\refi G.~I.~Taylor, \prsl 219 186 1953 ; R.~Aris, \prsl 235 67 1956 .
See also, J.~Koplik, S.~Redner, and D.~Wilkinson, \pra 37 2619 1988 .

\refi E. Ben-Naim, private communication.

\refi This is the so-called Matheron de Marsily model which was
introduced in G. Matheron and G. de Marsily, \wrr 16 901 1980 .  For
related work, see, \eg, J.-P. Bouchaud, A.~Georges, J.~Koplik,
A.~Provata, and S.~Redner, \prl 64 2503 1990 ; S. Redner, \pA 168 551
1990 .

\refi P.~S.~Hagan, C.~R.~Doering and C.~D.~Levermore, \jsp 54 1321 1989 ;
See also, C.~R.~Doering, P.~S.~Hagan and C.~D.~Levermore,
\prl 59 2129 1987 , and P.~S.~Hagan, C.~R.~Doering and C.~D.~Levermore,
\SIAMjam 49 1480 1989 .


\vfill\eject

\centerline{\bf Figure Captions}\smallskip

\rfigi (a) Wind shear in two dimensions with a ``cliff'' at $x=0$.
The underlying lattice (thin lines) is defined with $y=0$ bisecting two
horizontal rows of points.  A typical particle trajectory in this system
is sketched.  If the particle strays below the curve $y\cong-\sqrt{Dx/v}$,
it is likely to be absorbed.  (b) The same trajectory plotted with $x$
as a function of $t$.  The distribution of segment lengths $t_i$ in this
representation is proportional to $t^{-3/2}$ (see text).

\rfigi Time dependence of $S(t)$, based on exact enumeration of the
spatial probability distribution for two ``semi''-particles (each with
weight 1/2) initially at $(x_0,y_0)=(16,\pm 1/2)$, in a planar
semi-infinite system with $v=1$ and anisotropic diffusion with $D_y=1$,
$D_y=1/4$, and $D_y=1/16$ (upper to lower data sets).  The dashed
straight line has slope $-1/4$.

\rfigi Time dependence of $\av{x(t)^n}^{1/n}$ and $\av{y(t)^n}^{1/n}$
for $n=1$, 2 for the distribution of surviving walks for the case
$v=D_y=1$ and $x_0=1$.  The straight lines have slopes of 1 and
1/2.

\rfigi (a) Representative results for the time dependence of the
$S(t)$ based on exact enumeration of the probability distribution for a
planar semi-infinite system with isotropic diffusion and (i)
$(v,x_0,D)=(0.2,1,0.25)$, or $\pe=0.8$, (ii) $(0.1,1,0.25)$, $\pe=0.4$,
(iii) $(0.05,1,0.25)$ or $\pe=0.2$ (lower to upper data sets,
respectively).  Plotted is $S(t)\,\pe^{-3/4}$ versus $\tilde\tau$.  The
dashed line has slope $-1/4$.  (b) Schematic behavior for $S(t)$ versus
$t$ on a double logarithmic scale with isotropic diffusion in the limit
of $\pe\ll 1$.

\rfigi Schematic behavior for $S(t)$ versus $t$ on a double logarithmic
scale for a finite strip of width $w$ with anisotropic diffusion in the
limits of (a) $x_0\gg\ell$ and (b) $x_0\ll\ell$.

\vfill\eject\bye